\documentclass[12pt,preprint]{aastex}
\usepackage{amssymb}
\newcommand{\lperp}{\lambda_\perp}
\newcommand{\lpara}{\lambda_\parallel}
\newcommand{\Lpara}{\Lambda_\parallel}
\newcommand{\lmhd}{L_{\rm{mhd}}}

\newcommand{\kpara}{k_\parallel}
\newcommand{\vperp}{v_{\lambda_\perp}}
\newcommand{\vpara}{v_{\lambda_\parallel}}
\newcommand{\va}{v_{\rm{A}}}
\newcommand{\vstr}{v_{\rm{stream}}}
\newcommand{\alfven}{Alfv\'{e}n~}

\shorttitle{Wave damping by MHD turbulence}
\shortauthors{Farmer \& Goldreich}

\begin{document}
\title{Wave damping by MHD turbulence and its effect upon cosmic ray
  propagation in the ISM}
\author{Alison J. Farmer\altaffilmark{1} and Peter Goldreich\altaffilmark{1,2}}
\altaffiltext{1}{Theoretical Astrophysics, MC 130-33, Caltech,
  Pasadena, CA 91125, e-mail: ajf@tapir.caltech.edu, pmg@tapir.caltech.edu}
\altaffiltext{2}{SNS, IAS, Einstein Drive, Princeton, NJ 08540}

\begin{abstract}
Cosmic rays scatter off magnetic irregularities (\alfven waves) with
which they are resonant, that is waves of wavelength comparable to
their gyroradii. These waves may be generated either by the cosmic
rays themselves, if they stream faster than the \alfven speed, or by
sources of MHD turbulence. Waves excited by streaming cosmic rays are
ideally shaped for scattering, whereas the scattering efficiency of
MHD turbulence is severely diminished by its anisotropy. We show that
MHD turbulence has an indirect effect on cosmic ray propagation by
acting as a damping mechanism for cosmic ray generated waves. The hot
(``coronal'') phase of the interstellar medium is the best candidate
location for cosmic ray confinement by scattering from self-generated
waves. We relate the streaming velocity of cosmic rays to the rate of
turbulent dissipation in this medium, for the case in which turbulent damping
is the dominant damping mechanism. We conclude that cosmic rays
with up to $10^2\,$GeV could not stream much faster than the \alfven
speed, but that $10^{6}\,$GeV cosmic rays would stream unimpeded by
self-generated waves unless the coronal gas were remarkably
turbulence-free.
\end{abstract}

\section{Introduction}

Cosmic ray (CR) scattering by resonant \alfven waves has been proposed to
be essential to their acceleration by shocks (e.g. \citealt{bel78})
and to their confinement within the Galaxy (e.g. \citealt{kul69}).

Much of the interstellar medium (ISM) is thought to be turbulent,
providing a ready source of \alfven waves. However, MHD turbulence has
the property that, as energy cascades from large to small scales,
power concentrates in modes with increasingly transverse wavevectors,
i.e. perpendicular to the background magnetic field direction
(Goldreich \& Sridhar 1995, 1997). Cosmic rays scatter best off waves
that have little transverse variation, so CR scattering by MHD
turbulence is necessarily extremely weak, leading to very long CR mean
free paths (e.g. \citealt{cha00a}; \citealt{yan02}).

If cosmic rays stream faster than the \alfven speed, they can amplify
waves (naturally of the correct shape for scattering) through the
resonant streaming instability (see \citealt{wen74}). As the waves
amplify, the scattering strength increases and the streaming velocity
is reduced. For this process of self-confinement to operate, the
excitation rate of the waves by streaming cosmic rays must exceed the
sum of all rates of wave damping.

Wave damping depends upon the properties of the medium in which the
CRs propagate. Important mechanisms include ion-neutral collisions in
regions of partial ionization, and non-linear Landau damping in the
collisionless limit. In this paper we introduce another
mechanism, wave damping by background MHD turbulence. As cosmic ray-generated waves propagate along magnetic field lines, they are
distorted in collisions with oppositely-directed turbulent
wavepackets. As a result, the wave energy cascades to smaller scales
and is ultimately dissipated. This process, which is best viewed
geometrically, is described in \S~\ref{sec:damp}. MHD turbulence
thus becomes an impediment to the scattering of cosmic rays, as
opposed to just an ineffective scatterer of them. This mechanism was
mentioned briefly by \citet{cho02}, \citet{laz02} and \citet{yan02}.

The paper is arranged as follows. Relevant properties of the MHD
cascade are described in \S\ref{sec:mhd}, followed by an explanation
of the turbulent damping rate in \S\ref{sec:damp}. In
\S\ref{sec:competition} we describe the competition between growth and
damping of waves due to CR streaming. We apply these ideas to the
problem of Galactic CR self-confinement in \S\ref{sec:cr}, and use
this to place limits on the cascade rate of the turbulence in the
coronal gas, assuming that the observed streaming velocities are due to
self-confinement in this medium. In \S\ref{sec:other} we compare with
other work in this area and in \S\ref{sec:conc} we conclude.

\section{The MHD cascade as a damping mechanism}

\subsection{Relevant properties of the cascade}
\label{sec:mhd}

The strong incompressible MHD cascade proposed by Goldreich \& Sridhar
(1995, 1997) has the property that as the cascade proceeds to smaller
scales, power becomes increasingly concentrated in waves with
wavevectors almost perpendicular to the local mean magnetic field. We
envisage a situation in which turbulence is excited isotropically at
an MHD outer scale $\lmhd$ with RMS velocity fluctuations $v\sim\va$
and magnetic field fluctuations $\delta B\sim B_0$, where $B_0$ is the
magnitude of the background magnetic field.\footnote{Turbulence can
also be injected at smaller velocities on smaller scales, in which
case $\lmhd$ should be considered an extrapolation beyond the actual
outer scale of the cascade.} Well inside the cascade, the variations
parallel to the magnetic field are much more gradual than those
perpendicular to it, i.e. $v(${\boldmath{$\lambda$}}$) \simeq \vperp \gg \vpara$
if $\lperp \sim \lpara$.\footnote{Throughout this paper,
``perpendicular'' and ``parallel'' wavelengths refer respectively to
the inverse of the wavevector components perpendicular and parallel to
the background magnetic field direction.}  Equivalent relations hold
for magnetic field fluctuations. For fluctuations of a given
amplitude, therefore, the correlation length (defined so that $\vperp
\sim v_{\Lambda_\parallel}$) parallel to magnetic field lines,
$\Lpara$, is much greater than that perpendicular to them, $\lperp$.
Turbulent eddies are highly elongated parallel to magnetic field
lines. 

Strong MHD turbulence is characterized by ``critical balance''. In
other words, a wavepacket shears at a rate which is comparable to its
frequency $\omega =\va \kpara\sim \va/\Lpara$ and is also of order
$v_{\lambda_\perp}/\lambda_\perp$.  Thus
\begin{equation}\frac{\vperp}{\lperp} \sim \frac{\va}{\lpara} \, . 
\label{eq:crit}
\end{equation}

Application of the Kolmogorov argument for the constancy of the energy
cascade rate $\epsilon$ per unit mass yields
\begin{equation} \epsilon \sim \frac{v^2}{t_{\rm{cascade}}} \sim
  \frac{\vperp^3}{\lperp} \sim \frac{\va^3}{\lmhd},
\end{equation} 
from which we obtain the fluctuation amplitude on perpendicular
scale $\lperp$,
\begin{equation}
  \vperp\sim\va\left(\frac{\lperp}{\lmhd}\right)^{1/3}\sim (\epsilon
  \lperp)^{1/3}.
\label{eq:vperp}
\end{equation} 
An analogous relation holds for magnetic field perturbations. Well inside the cascade, $v \ll \va$ and $\delta B \ll
B_0$. We
combine equations (\ref{eq:crit}) and (\ref{eq:vperp}) to obtain the eddy shape:
\begin{equation}
\Lpara(\lperp) \sim \lmhd^{1/3} \lperp^{2/3} > \lperp\, .
\label{eq:lpara}
\end{equation}

\subsection{The turbulent damping rate}
\label{sec:damp}

The energy cascade from large to small scales in MHD turbulence is due
to distortions produced in collisions between oppositely-directed
\alfven wavepackets. This is best visualized geometrically as being due
to the shearing of wavepackets as they travel along wandering magnetic
field lines. A good description is given in \citet{lit01}.

Consider the fate of a wavepacket with initial perpendicular and
parallel wavelengths $\lperp$ and $\lpara$. It suffers an order unity
shear after traveling over a distance along which the fields lines
that guide it spread by order $\lambda_\perp$. By then the energy it
carries has cascaded to smaller scales, ultimately to be dissipated as
heat at the inner scale. This process occurs not only for waves that
are part of the turbulent cascade, but also for any other \alfven
waves in the medium. As these waves travel along the field lines, they
are distorted in collisions with oppositely-directed turbulent
wavepackets.

On a perpendicular scale $\lperp$, the field lines spread by order
unity over a parallel distance $\Lpara$, where $\Lpara(\lperp)$ is a
property of the background turbulence and is given by
equation (\ref{eq:lpara}). Therefore any wavepacket of perpendicular scale
$\lperp$ cascades once it travels this distance. Because of the nature
of the MHD cascade, this corresponds to many wave periods for a wave
with $\lpara \lesssim \lperp \ll \Lpara$ (but to one wave period for
waves shaped like those in the turbulent cascade, as described by
critical balance). The damping rate is a
function of $\lperp$:
\begin{equation}\Gamma_{\rm{turb}} \sim \frac{1}{t_{\rm{cascade}}(\lperp)} \sim
\frac{\vperp}{\lperp} \sim \frac{\va}{\lmhd^{1/3} \lperp^{2/3}}
\sim \frac{\epsilon^{1/3}}{\lperp^{2/3}}.\label{eq:damp}
\end{equation}
This damping rate applies to any wave with perpendicular wavelength
$\lperp$ propagating in a background of strong MHD turbulence, so long
as $\lmhd \gg \lperp \gg l_{\rm{dissipation}}$. The appropriate value
of $\lperp$ to use for CR-generated waves will be considered in
\S\ref{sec:grdamp}.

\section{Competition between growth and damping}
\label{sec:competition}
\subsection{Resonant scattering of cosmic rays}
As cosmic rays stream along magnetic field lines, they are scattered
in pitch angle by magnetic irregularities (\alfven waves, of
appropriate shape; see below), and thus exchange momentum (and energy)
with particular waves. If cosmic rays stream faster than the \alfven
speed, they can excite \alfven waves traveling in the same
direction. Provided the excitation rate exceeds the total damping rate
due to other processes, the waves amplify exponentially. Initial
perturbations too weak to significantly scatter CRs can strengthen
until the scattering reduces the CR streaming velocity. Even thermal
fluctuations could provide seed waves in the absence of other
sources. The reduction of the streaming velocity by cosmic ray-amplified waves is known as self-confinement.  Next we describe which
random fluctuations are selectively amplified by cosmic ray protons
with energy $\gamma$ GeV.

\subsubsection{Parallel lengthscale}

Cosmic rays spiraling along a mean magnetic field $\mathbf{B_0}=B_0
\mathbf{\hat{z}}$ scatter in pitch angle off \alfven waves with which
they are parallel-resonant, i.e. waves for which
\begin{equation}
\kpara=\frac{1}{\mu r_L},
\end{equation} 
where $\mu$ is
the cosine of the particle's pitch angle, and $r_L$ is its gyroradius
(\citealt{kul69}; \citealt{wen74}). On the timescale of the CR's
passage, the wave is almost static since the CR is relativistic and
$\va \ll c$.  Thus the wave's time dependence is neglected in the
above resonance condition. When resonance holds, the cosmic ray
experiences a steady direction-changing force.

\subsubsection{Perpendicular lengthscale}

A cosmic ray is most efficiently scattered by parallel-propagating
waves, $\lperp \gg \lpara \sim r_L$, because in these the direction
changing force maintains a steady direction in one gyroperiod. Moving
through waves that have significant perpendicular components, $\lperp \ll
\lpara$, the cosmic ray traverses many perpendicular wavelengths,
leading to oscillations of the direction-changing force and
inefficient scattering. This explains why cosmic rays are weakly
scattered by MHD turbulence (see e.g. Chandran 2000a; \citealt{yan02}),
and also why the waves in the turbulent cascade damp faster than
cosmic rays can excite them.


The closer to parallel waves propagate, the faster streaming cosmic
rays can excite them. The growth rate for waves that are
parallel-resonant and reasonably close to parallel-propagating
($\lpara \lesssim \lperp$) is given by (see \citealt{kul69})
\begin{equation}
\Gamma_{\rm{cr}}(\kpara) \sim \Omega_0
\frac{n_{\rm{cr}}(>\gamma)}{n_{\rm{i}}}
\left(\frac{\vstr}{\va}-1\right),\label{eq:crgrowth}
\end{equation}
where $\vstr$ is the net streaming velocity of the cosmic rays
measured in the rest frame of the ISM, $\Omega_0=eB_0/mc$ is the CR
cyclotron frequency in the mean field, $n_{\rm{i}}$ is the ion number
density in the ISM, and $n_{\rm{cr}}(>\gamma)$ is the number density
of cosmic rays with gyroradius $r_L > \gamma mc^2/eB_0=1/\kpara$,
i.e. those particles which can, for the appropriate value of $\mu$, be resonant with waves of parallel
wavevector $\kpara$. Because the cosmic ray energy spectrum is steep,
the energies of most resonant particles are close to the lowest energy
that permits resonance with the wave. Therefore we associate $\kpara
\sim 1/r_L(\gamma)$ and $n_{\rm{cr}}(> \gamma) \simeq \gamma
n_{\rm{cr}}(\gamma)$.\footnote{Particles with close to 90-degree pitch
  angles ($\mu \ll 1$)
are scattered mainly by mirror interactions \citep{fel01}.}

\subsection{Growth and damping}
\label{sec:grdamp}

Growth rates are highest, and damping rates lowest, for the most
closely parallel-propagating waves, that is, those waves with largest
$\lperp$. Therefore, we consider the limiting case of the most
parallel-propagating wave that can be excited. This most-parallel wave
sets the minimum streaming velocity required for the instability to
operate. The limit to parallel propagation is set by the turbulent
background magnetic field: the largest wave aspect ratio possible is
fixed by the straightness of the field lines. In the presence of MHD
turbulence, the field direction depends on position. The change in
direction across scale $\lperp$ is set by turbulent field fluctuations
on this scale. It is not meaningful to talk about waves propagating at
an angle less than $\delta B(\lperp)/B_0$ away from parallel, because
the field direction changes by this much across the wavepacket. We can
therefore only have waves with
\begin{equation}
\frac{\lpara}{\lperp} > \frac{\delta B(\lperp)}{B_0} \sim
\left(\frac{\lperp}{\lmhd}\right)^{1/3} \sim \left(\frac{\epsilon
  r_L}{\va^3}\right)^{1/4}, 
\label{eq:minkperp}\end{equation}
where we have used $\lpara \sim r_L$, the resonance condition.

To obtain the damping rate of the most closely parallel-propagating
wave, we substitute equation (\ref{eq:minkperp}) into equation
(\ref{eq:damp}), which yields
\begin{equation}
\Gamma_{\rm{turb,min}} \sim \left(\frac{\epsilon}{r_L \va}\right)^{1/2};
\label{eq:tdamp}
\end{equation} 
all other waves damp faster than this one, for given $r_L$.

We can view the damping as being due to the introduction of
perpendicular wavevector components to the CR-generated wave. This is
how the background turbulence cascades, and the CR-generated wave is
being integrated into the cascade. We can decompose the modified wave
into components with almost-perpendicular and almost-parallel
wavevectors. The perpendicular part is not excited and is more
strongly damped, but the almost-parallel-propagating component
continues to be amplified by resonant cosmic rays. 

For the instability to
operate, we require the maximum possible growth rate
(eq. [\ref{eq:crgrowth}]) to be larger than the minimum damping rate
(eq. [\ref{eq:tdamp}]):
\begin{equation}
\Gamma_{\rm{cr}}[F_{\rm A}(\gamma)] > \Gamma_{\rm{turb,min}}(\gamma).
\label{eq:balance}
\end{equation}
where $F_{\rm A}(\gamma)\sim (\vstr-\va)n_{\rm{cr}}(>\gamma)$ is the
cosmic ray flux measured in the frame moving with the waves. Equation
(\ref{eq:balance}) can be written in the form $F_{\rm A}(\gamma) >
F_{\rm crit}(\gamma)$.  If $F_{\rm A}$ is less than $F_{\rm crit}$
then wave amplification does not occur and the cosmic rays are not
significantly scattered. Equivalently, the resonant streaming
instability cannot reduce $F_{\rm A}$ below $F_{\rm crit}$.

If the instability is to confine cosmic rays to regions of shock
acceleration, or to the Galaxy (which we discuss in the next section),
then the level of background turbulence must be low enough to permit
the growth of resonant waves.

\section{Application to cosmic ray self-confinement in the ISM}
\label{sec:cr}

Cosmic rays are preferentially produced in the denser regions of the
Galaxy, and they escape from its edges. Two lines of evidence imply
that they do not stream freely out of the Galaxy: the CR flux in the
solar neighborhood is observed to be isotropic to within $\sim$0.1\%
at energies less than $\sim 10^6$ GeV, and the abundance of the
unstable nucleus $^{10}$Be produced by spallation establishes that CRs
are confined within the Galaxy for $\sim 10^7$ years \citep{sch02}.

Scattering by \alfven waves has been viewed as the leading mechanism
for confinement. Both waves associated with background MHD turbulence
and those resonantly excited by cosmic rays have been considered in
this regard. Prior to the recognition that MHD turbulence is
anisotropic, the former were generally favored. Now self-confinement
appears to be the more viable option.

The most promising location for the operation of the streaming instability is the hot ISM (HISM), a.k.a. the coronal gas
(\citealt{ces81,fel01}). The abundances of cosmic ray nuclei produced
by spallation suggest that cosmic rays spend about two-thirds of their
time in this medium (see e.g. \citealt{sch02}). Ion-neutral damping of
waves is ineffective in the HISM. The coronal gas is hot ($T\sim10^6$
K) and tenuous ($n_{\rm{i}} \sim 10^{-3} \rm{~cm}^{-3}$), with an
\alfven velocity, assuming $B_0\sim 3~\mu\rm{G}$, of $\va\sim 2 \times
10^7 \rm{~cm~s}^{-1}$. The gyroradius of a relativistic proton in this
field, $r_L \sim 10^{12} \gamma\rm{~cm}$, lies within the inertial
range of the MHD cascade.

Assuming the cosmic ray density in the HISM to be similar to that near
the Sun,\footnote{If $n_{\rm cr}$ is lower than it is near the Sun,
then confinement will begin to be problematic at lower energies, and
vice versa} $n_{\rm{cr}}(> \gamma) \simeq 2 \times 10^{-10}
\gamma^{-1.6}\rm{~cm}^{-3}$ \citep{wen74}, we can calculate the
velocity above which the streaming instability in the HISM would turn
on, assuming our turbulent damping to be the dominant damping mechanism. To accomplish this, we substitute equations (\ref{eq:crgrowth}) and
(\ref{eq:tdamp}) into inequality (\ref{eq:balance}), treating it
as an equality. We find
\begin{eqnarray}
\vstr &\sim&
\va\left[1+\frac{n_{\rm{i}}}{n_{\rm{cr}}(\gamma)}\frac{\omega_0}{\Omega_0}
\left(\frac{\lmhd}{r_L}\right)^{1/2}\right]\nonumber\\ &\sim
&\va\left[1+ \left(\frac{\epsilon}{700
\rm{~erg~s}^{-1}\rm{~g}^{-1}}\right)^{1/2} \gamma^{1.1} \right]\,\, ,
\label{eq:vstr}
\end{eqnarray}
where $\omega_0 = \va/\lmhd$ is the turbulent decay rate on the outer
scale.

The mean rate at which turbulent dissipation heats the coronal gas is
unlikely to exceed its radiative cooling rate, $\epsilon \sim 0.06
\rm{~erg~g}^{-1}\rm{~s}^{-1}$ for solar abundances
(\citealt{bt87}, p580). Unfortunately, we do not know whether the heating is
continuous or episodic, and what fraction is due to shocks as opposed
to turbulence.\footnote{It seems plausible that shocks, especially
if they intersect, would efficiently excite turbulence.}

Roughly one supernova explosion occurs per century in the Galaxy, or
on average one per square 100 pc of the disk every $1\times 10^6
\rm{~yr}$. Turbulence injected with $v\sim\va$ on scales $L\sim
100\rm{~pc}$ decays in a time $L/\va \sim 5\times 10^{5}\rm{~yr}$, so
it might be replenished before decaying. However, supernovae occur
predominantly in the Galactic plane and it is uncertain how effective
they are in stirring the coronal gas, which has a large vertical scale
height. Suppose that each supernova releases $10^{51}\rm{~erg}$ of
mechanical energy that is ultimately dissipated by turbulence. This
amounts to a dissipation rate of $3 \times 10^{41}\rm{~erg~s}^{-1}$
which, if evenly distributed by volume throughout a disk of radius 10
kpc and thickness 1 kpc, would provide a mean heating rate of
$\bar{\epsilon} \sim 25 \rm{~erg~s}^{-1}\rm{~g}^{-1}$ in the
HISM. This value is much greater than our estimate of the radiative
cooling rate.

The cosmic ray anisotropy measured locally is $\lesssim 0.1\%$ for
$\gamma \lesssim 10^6$ \citep{sch02}, i.e. up to the ``knee'' in the
CR energy spectrum. The \alfven velocity in the HISM is of the same
order as the local streaming velocity: $\va/c \simeq 0.1\%$. 
Substituting into equation (\ref{eq:vstr}) the value of $\epsilon$
obtained by balancing the radiative cooling of the hot gas with
heating due to steady state turbulent dissipation, we obtain
\begin{equation}
\vstr \sim \va(1+9 \times 10^{-3}\gamma^{1.1}).
\label{eq:vstr2}
\end{equation}
Equation (\ref{eq:vstr2}) suggests that self-confinement in the HISM might
account for the small observed cosmic ray anisotropy up to $\gamma \lesssim
10^2$, but not much beyond. To limit the streaming velocity of protons
with $\gamma\sim 10^6$ to $\sim \va$ would require the turbulent dissipation rate
to be astonishingly low, $\epsilon \lesssim 4 \times
10^{-11}\rm{~erg~s}^{-1}\rm{~g}^{-1}$. 

\subsection{Comparison with previous work}
\label{sec:other}

That background MHD turbulence might be an impediment to the
self-confinement of cosmic rays was mentioned briefly in
Lazarian et al. (2002), \citet{yan02} and Cho et al. (2003).

\citet{kul78} proposed non-linear Landau damping as the dominant
damping mechanism for cosmic ray-generated waves in the HISM. Wave
damping occurs when plasma ions ``surf'' on beat waves produced by the
superposition of CR-generated waves. The damping rate for this
process,\footnote{We use the unsaturated damping rate, as justified in
\citet{fel01}.} for similar HISM parameters as adopted in this paper,
gives $\vstr \simeq \va(1 + 0.05 \gamma^{0.85})$ \citep{ces81}. This
predicted streaming velocity is not very different from that obtained
in equation (\ref{eq:vstr2}). Both damping mechanisms are too strong
to permit self-confinement to reduce the streaming velocity of high
energy cosmic rays to the locally observed levels.

\citet{cha00b} proposes that magnetic mirror interactions in dense
molecular clouds may provide confinement of high energy cosmic
rays. The present paper provides further support for the idea that
a confinement mechanism other than scattering by \alfven waves is
dominant for high energy cosmic rays.

\section{Conclusions}
\label{sec:conc}
A background of anisotropic MHD turbulence acts as a linear damping
mechanism for MHD waves excited by the streaming of cosmic rays.  Low
energy cosmic rays are numerous enough to excite \alfven waves in the HISM
when streaming at velocities compatible with observational limits on their
anisotropy. However, high energy Galactic cosmic rays could only be
self-confined to stream this slowly if turbulent dissipation in the HISM accounted for only
a tiny fraction of its heat input.

\acknowledgments{
This research was supported in part by NSF grant AST 00-98301. The authors thank Russell
Kulsrud for raising the issue discussed in this paper and for illuminating discussions.
}

\end{document}